\documentstyle[aps]{revtex} 
\newcommand{\be}{\begin{equation}} 
\newcommand{\ee}{\end{equation}} 

\begin{document}

\title{Studies of Phase Turbulence 
in the One Dimensional Complex Ginzburg-Landau Equation}

\author{Alessandro Torcini, Helge Frauenkron and Peter Grassberger}

\address{\it Theoretical Physics,
Wuppertal University, D-42097 Wuppertal, Germany}

\maketitle

\vspace{2.cm}
\centerline{\today}

\begin{abstract}
The phase-turbulent (PT) regime for the one dimensional complex
Ginzburg-Landau equation (CGLE) is carefully studied, in the limit 
of large systems and long integration times, using an efficient
new integration scheme. Particular attention is paid to solutions 
with a non-zero phase gradient. For fixed control parameters, 
solutions with conserved average phase gradient $\nu$ exist only 
for $|\nu|$ less than some upper limit. The transition from phase 
to defect-turbulence happens when this limit becomes zero. A 
Lyapunov analysis shows that the system becomes less and 
less chaotic for increasing values of the phase gradient. 
For high values of the phase gradient a family of non-chaotic 
solutions of the CGLE is found. These solutions consist of 
spatially periodic or aperiodic waves travelling with constant 
velocity. They typically 
have incommensurate velocities for phase and amplitude propagation, 
showing thereby a novel type of quasiperiodic behavior. The main 
features of these travelling wave solutions can be explained through 
a modified Kuramoto-Sivashinsky equation that rules the 
phase dynamics of the CGLE in the PT phase. The latter explains also 
the behavior of the maximal Lyapunov exponents of chaotic solutions.

\end{abstract}

\pacs{PACS numbers: 47.27.Cn, 05.45.+b, 47.27.Eq}

\section{Introduction}

The complex Ginzburg-Landau equation (CGLE) plays a fundamental r\^ole 
in the study of spatially extended systems. It describes the dynamics of 
a generic spatially extended system which undergoes a Hopf bifurcation 
from a stationary to an oscillatory state. Sufficiently close to the 
bifurcation point, any such system can be described by the CGLE 
\cite{kura_book}. For a review of experimental systems described 
by the CGLE see \cite{exp}.

In the present paper we want to discuss only the CGLE in one spatial 
dimension. Even though this lacks important features of the CLGE 
in higher dimensions, it displays a variety of chaotic and stationary 
regimes for different values of control parameters.
Recently, Shraiman {\it et al.} \cite{shra} and Chat{\'e} \cite{chate}
have provided a careful description of these regimes.
In particular, four different chaotic states have been identified:
the phase-turbulent (PT), the defect-turbulent (DT), the bichaotic
and the spatiotemporal intermittent regime \cite{chate}. The 
fundamental states are the PT and the DT regimes. The bichaotic state 
is characterized by the fact that PT or DT chaos can arise, depending 
on the initial conditions --- while DT chaos or stable (periodic) 
solutions can appear in the intermittent region. The PT and DT phases 
have been the object of several detailed studies 
\cite{shra,kura,saga,green}. 

Let us write the one-dimensional CGLE as
\be
   A_t = (1+ic_1) A_{xx} + A -(1-ic_3)|A|^2A               \label{cgle}
\ee
where the parameters $c_1$ and $c_3$ are real positive numbers,
while $A(x,t)=\rho(x,t) \exp[i \psi(x,t) ]$ is a complex field
of amplitude $\rho$ and phase $\psi$. Eq.~(\ref{cgle}) reduces 
in the limit $(c_1,c_3) \to (\infty,\infty)$
to the integrable nonlinear Schr{\"o}dinger equation, an
equation well known to admit soliton-like solutions \cite{book_nlse}. 
In the opposite limit, $(c_1,c_3) \to (0,0)$, the
real Ginzburg-Landau equation will be recovered, an equation related
to symmetry-breaking instabilities of non-oscillatory type
\cite{kura_book}. 

Equation (1) admits plane-wave solutions of the form
\be
    A_0(x,t)= \sqrt{1 - k^2} \exp[i (k x - \Omega_0 t)]     \label{plane}
\ee
where $\Omega_0 = -c_3+(c_1+c_3) k^2$. Below the so-called 
Benjamin-Feir (BF) line defined by $c_3=1/c_1$ these solutions are 
linearly stable provided $k^2 <(1-c_1 c_3)/(2(1+c_3^2)+1-c_1c_3)$ 
\cite{mann}. Outside this ``Benjamin-Feir band'', one has an Eckhaus 
instability against long-wavelength modes. Slightly below the BF line one 
finds multistability, i.e. the final solution depends on the initial 
conditions \cite{chate}. Above it, i.e. for $c_3c_1>1$, all plane wave 
modes are unstable. The latter is the region we are interested in.

For generic (not necessarily periodic) solutions, the {\it average} 
phase gradient is defined as 
\be
    \nu = {1 \over L} \int_0^L dx \partial_x \psi(x,t) \,,    \label{wind}
\ee
where $L$ is the system size (we assume periodic boundary conditions 
$A(x+L,t)=A(x,t)$ throughout this paper, unless stated differently). 
The PT phase is defined as 
that regime where $\nu$ is conserved \cite{kura}. It is encountered 
just above the BF line (i.e. for $c_3 > 1/c_1$). It is a state where the 
chaotic behavior of the field is essentially ruled by the phase 
dynamics. The amplitude, always bounded away from zero for Eq.~(\ref{wind}) 
to make sense, shows small fluctuations. 

This is different in the DT regime in which the amplitude-dynamics 
becomes essential \cite{shra,kura,saga}. Large amplitude  oscillations 
are observed in this phase which occasionally drive $\rho(x,t)$ down to 
zero. Events where this happens are called defects. At a defect, the 
phase is of course no longer defined, and the appearance of a defect 
implies that the phase difference $\psi(x+\epsilon,t)-\psi(x-\epsilon,t)$ 
jumps by $\pm 2\pi$. Thus $\nu$ is no longer conserved. 

The transition from the PT to the DT regime happens at a line $L_1$ in 
the ($c_1,c_3$)-plane. To locate this transition precisely 
\cite{shra,green}, one needs some order parameter which allows to 
distinguish the two phases. Several parameters have been proposed 
recently: the density of defects, the phase and amplitude correlations 
lengths, and the Kaplan-Yorke dimension. But as observed in \cite{green}, 
the latter are not suitable as order parameters as they do not show 
any non-analytic behavior at the transition line $L_1$. The only good 
order parameter among the above candidates is the defect density 
$\delta_D$. Its value is $> 0$ in the DT regime, while it vanishes when
approaching the PT phase \cite{shra,green}. To present other suitable 
order parameters will be one of the aims of the present paper.

In the PT regime, all observables depend of course on the value of $\nu$, 
as ergodicity is broken in this phase. Nevertheless, in literature only 
few studies have been devoted to solutions with $\nu = 0$ 
\cite{baleari,io}. To fill this gap is the second goal of the present 
paper. It will be shown that most observables depend indeed rather 
strongly and systematically on $\nu$.

Since our study is mostly numerical, we need an efficient integration 
routine. We found that a novel scheme, similar to but more efficient than 
the one introduced in \cite{torc1}, gave excellent results. This scheme 
is introduced in the next section and compared with the usual
pseudo-spectral codes. In Section III the transition from phase to defect
turbulence is investigated with the help of a new order parameter. A 
Lyapunov analysis for the solutions with $\nu \ne 0$ is reported 
in Section IV together with a detailed characterization of the 
observed stable solutions. In Section V it is shown how a modified 
Kuramoto-Sivashinsky equation for the phase is able to reproduce
the main features of the dynamics of the considered solutions of the CGLE. 
Our results are summerized in Sect. VI.

\section{Integration Scheme}

The algorithm applied in this work is a operator splitting scheme similar 
to the well known ``leap frog'' algorithm for ordinary differential equations 
(this should not be confused with the ``staggered leap frog'' \cite{numeric} 
for PDE's). This algorithm is based on the fact that the dynamics consists 
of two independent terms, one nonlinear but local and the other nonlocal 
but linear, each of which by itself can be easily integrated. In the 
present case, this splitting is chosen as 
\be
   {\partial A \over \partial t} = {\cal N}A + {\cal L}A      \label{pde}
\ee
with 
\be
   {\cal N}A = A -(1-ic_3)|A|^2A \quad , \quad 
   {\cal L}A = (1+ic_1){\partial^2 A \over \partial x^2} \;.
\ee
For a small time step $\tau$, the formal solution $A(t) = 
e^{({\cal N}+{\cal L})t} A(0)$ 
of Eq.~(\ref{pde}) is then approximated by means of the Trotter formula 
\be
   e^{({\cal N}+{\cal L})\tau} = e^{{\cal N}\tau/2}
                 e^{{\cal L}t}e^{{\cal N}\tau/2} + O(\tau^3) \;. 
                                                  \label{trotter}
\ee
For a finite time $t=n\tau, \;n>>1$, we can lump together $n-1$ half steps 
and arrive at 
\be
   e^{({\cal N}+{\cal L})t} \approx e^{{\cal N}\tau/2} [e^{{\cal L}t}
         e^{{\cal N}\tau}]^{n-1}e^{{\cal L}t}e^{{\cal N}\tau/2}\;.
\ee
Notice that we refrain from giving here a formal error estimate. For a 
chaotic system, the error will of course increase exponentially, and 
the exponent will depend on the maximal Lyapunov exponent $\lambda$: 
roughly we expect $\delta A(t) \sim e^{c\lambda\tau^2 t}$, where $c$ is 
the constant in the $O(\tau^3)$ correction in Eq.~(\ref{trotter}).

Solving the nonlinear evolution $e^{{\cal N}\tau}$ is easy, by considering 
amplitude and phase separately. We first find 
\be 
   \rho_1(x,t+\tau) \equiv [e^{{\cal N}\tau}\rho](x,t) = 
       \left[ e^{-2 \tau} (1/\rho(x,t)^2-1)+1 \right]^{-1/2} \;.
                                                      \label{rhosol}
\ee
Inserting this into the equation for the phase, the latter can also be 
solved exactly to obtain
\be
   \psi_1(x,t+\tau) \equiv [e^{{\cal N}\tau}\psi](x,t) = \psi(x,0) + c_3
       \{\tau + \log [\rho(x,t) / {\rho_1} (x,t+\tau) ]\}
                                                      \label{phasol}
\ee
A non-trivial problem appears only for the solution $e^{{\cal L}t}$ 
of the linear part since this is non-local. It is only here that 
our algorithm deviates from previous applications of the leap frog 
algorithm to PDE's \cite{bandrauk,takahashi,mclachlan,frauen1,frauen2}. 
In these papers, the solution of the nonlocal part was performed in Fourier 
space. This means that two Fourier transformations have to be performed 
for each integration step. It leads to what will be called pseudospectral 
method in the following. In spite of the speed of FFT algorithms, it 
can be quite time consuming. It can also lead to intolerable round-off 
errors, which was the main reason in \cite{torc1} to perform this step 
by integrating with the Greens function of ${\cal L}$. More precisely, let 
us denote by $K(x,t)$ the solution of 
\be
   \partial K /\partial t = {\cal L} K
\ee
with initial condition $K(x,0) = \delta(x)$. For the present case it is 
simply
\be
   K(x,\tau) = \sqrt{{{\beta} \over \pi}} e^{- \beta_r x^2}
      \left[ \cos(\beta_i x^2) - i \sin(\beta_i x^2) \right] 
                                                           \label{kernel}
\ee
with $\beta=\beta_r-i\beta_i= [4 \tau(1+ic_1)]^{-1}$. Then 
\be
   [e^{{\cal L}\tau}A](x,t) = \int d\xi K(\xi,\tau) A(x-\xi,t).
                                                           \label{convol}
\ee
Since $K(\xi,\tau)$ can be computed numerically and stored once at the 
beginning, the algorithm essentially involves one convolution for each 
time step.

In practice, space has of course to be discretized as well. Let us 
denote $A_i(t) = A(i\Delta,t)$, where $\Delta$ is the resolution 
of the spatial grid. The most straightforward approach (used indeed in 
\cite{torc1}) would seem to start from Eq.~(\ref{convol}), 
and replace simply the integral by a finite sum over the central site 
plus its $2N$ nearest neighbors, which would lead to 
\be
   [e^{{\cal L}\tau}A]_i(t) \approx 
         \sum_{j=-N}^N K(j\Delta,\tau) A_{i-j}(t) \;.  \label{discrete1}
\ee
But this is not the optimal discrete approximation to Eq.~(\ref{convol}).
To obtain a more precise one, we first notice that since $e^{{\cal L}\tau}A$ 
is a linear operator symmetric under translations and reflections 
$x\to -x$, we can write any lattice approximation of it as 
\be
   [e^{{\cal L}\tau}A]_i(t) \approx \sum_{j=0}^N K_j (A_{i-j}(t)+A_{i+j}(t))
                                                        \label{discrete2}
\ee
with unknown coefficients $K_j$. Next we observe that we can decompose 
$A(x,t)$ into Fourier modes. Let us assume that a good approximation 
is obtained by a superposition of $2N+1$ modes $e^{ik\alpha x}$ with 
$k=-N,\ldots N$, and $\alpha$ some yet unspecified positive constant.
We then can determine the $K_j$ such that it gives the exact evolution 
on these modes. A straightforward calculation shows that this implies
\be
   K_0+2 \sum_{j=1}^{N} K_j \cos( j \Delta k \alpha) = e^{-(1+ i c_1)t
           k^2 \alpha^2} \qquad \forall k=0, \dots , N \;.
                                                  \label{conditions}
\ee
Finally, the value of $\alpha$ is chosen such that the $L_2$ norm 
$\{|K_0|^2+2\sum_{j>0}|K_j|^2\}^{1/2}$ of $K$ is minimal. With this 
choice, we hopefully minimize the effect of the neglected modes.

In the limit $N\to\infty$, the optimal set of Fourier modes are those 
with $\alpha = 2\pi/\Delta$, and the above scheme should give identical 
results (up to round-off errors) as the pseudospectral method obtained 
by applying a discrete FFT to $A_t={\cal L}A$ and integrating it in 
Fourier space. Thus the error committed in our method with finite $N$ 
cannot be smaller than the error committed by the FFT method, and it 
should tends towards it for $N\to\infty$.

In order to compare the precision of our algorithm with that of the 
pseudospectral code, we have evaluated the mean square errors 
accumulated during a fixed total integration time $T=1$ for 
different values of the spatial resolution $\Delta$ and of the number 
$N$ of convolution channels. In all simulations reported below, 
the time step and the parameter of the CGLE have been $\tau=0.05$, 
$c_1=3.5$ and $c_3=0.9$. The errors have been estimated by measuring the 
distance (in $L_2$ norm) between an orbit obtained by the algorithm 
to be tested (`test orbit') and a reference orbit obtained by the 
pseudospectral method. In order to reduce statistical fluctuations, 
the reference orbit was integrated for a time $>>T$, the test 
orbit was synchronized to it periodically at times $t=mT, \;m=1,2,\ldots$, 
and the distances building up during the intervals $mT<t<(m+1)T$ were 
averaged over.

For estimating the error due to time discretization, we used a 
reference orbit with the same $\Delta$ as the test orbit, but with 
four times smaller $\tau$. To estimate the space discretization error, 
the reference orbit had the same $\tau$, but half the value of $\Delta$ 
(and, in effect, $N=\infty$). Results are shown in Fig.~1. From these 
data we can draw the following conclusions: \\
(i) For all studied 
values of $\tau$ and $\Delta$, the time discretization had a much 
bigger effect than space discretization. We could of course have used 
a smaller value of $\tau$ (or a larger $\Delta$) to avoid this, but our 
choices were motivated by previous simulations \cite{shra,chate,green}. 
Nevertheless, we propose that future simulations should use finer 
time and/or coarser space grids. \\
(ii) The convergence of the spatial truncation errors with $N$ is 
quite fast. Typically, the error is less than twice the error of the 
FFT code for $N\geq 20$. This is in contrast to the scheme of 
\cite{torc1}, where $N$ had to be chosen larger than $\approx 60$. 
Due to the large time discretization error, we could indeed safely use 
$N=10$.\\
(iii) The FFT code needs a CPU time which increases logarithmically 
with the number $L/\Delta$ of mesh points, while our algorithm is 
essentially linear in $N$. Using the FFT routine C06ECF of the 
NAG library, the ratio between the CPU times needed by our algorithm 
and the pseudospectral one was $0.24N/\ln(L/\Delta)$. For typical 
simulations reported in this paper ($N=10, L=1024 - 4096, \Delta = 
1/2$), this ratio is $\approx 0.25 - 0.3$. Thus our algorithm is indeed 
considerably faster than the pseudospectral one.

\section{Order Parameters and the PT-DT Transition}

To obtain configurations with non-vanishing $\nu$, we used several 
different initial conditions. A first choice was 
\be
   A_k(t=0) = e^{i\nu k \Delta} + (r_k+ ip_k)               \label{in1}
\ee
with $r_k$ and $p_k$ random numbers uniformly distributed in $[-0.1,0.1]$. 
Notice that the randomness is needed to break spatial translation 
invariance. Another ansatz was
\be
   A_k(t=0) = e^{i\nu k\Delta+s_k}\rho_k(0)                 \label{in2}
\ee
where $\rho_k(0)$ was uniformly distributed in $[0.2,1.2]$, and 
$s_k \in [-0.01,0.01]$. Finally, in a third type of initial conditions 
we introduced also long range correlations in the amplitude, 
\be
   \rho_k(0) = 0.95\,\rho_{k-1}(0)+q_k, \quad q_k \in [-0.05,0.05]
\ee

All these initial conditions give consistent final results: in the regime 
identified as phase turbulent by previous authors, $\nu$ is indeed
conserved for sufficiently small initial values. However, for each pair 
$(c_1,c_3)$ there exists an upper limit $\nu_M$ above which defects arise 
in the system, leading finally to a phase gradient $\nu \le \nu_M$. 
Therefore, defects will arise also in the PT regime, provided that the 
value of $\nu$ is sufficiently high. The distribution $P(\rho)$ of the 
amplitude in the statistically stationary state becomes in general wider 
with increasing $\nu$, as shown in Fig.~2. In particular, the lower limit 
$\rho_{min}$ of its support decreases with $\nu$. But it seems that 
this decrease in discontinuous, so that $\rho_{min}$ cannot be used 
as a good order parameter. We will come back to this point later.

The dependence of $\nu_M$ on $c_3$ is plotted in Fig.~3 for $c_1 = 3.5$. 
This fixed value of $c_1$ was chosen for ease of comparison with previous 
studies. We see an almost linear decrease, and $\nu_M$ reaches zero 
exactly at the line $L_1$ which marks the onset of DT (this was also 
seen in \cite{baleari}). 

Thus $\nu_M$ can be used as order parameters to 
characterize the transition from phase to amplitude turbulence. From the 
general theory of continuous phase transitions we would in general not 
expect a linear behavior of the order parameter as the transition is 
approached. Thus we should be suspicious that the linearity seen in Fig.~3 
might give way to a power law for very small $\nu_M$. 

In order to understand this better and to characterize better the 
transition between phase and defect turbulence (always concentrating 
on the value $c_1=3.5$ which we assume to be a generic point on the 
line $L_1$), we performed several simulations with high statistics.  
For each value of $c_3$ we performed $\sim 50$ to 70 simulations with 
$\nu>\nu_M$ and with independent initial conditions, and followed them for 
$\geq 10,000$ time units (some simulations were followed even for 
$>10^5$ time units), unless a defect is found --- in which case $\nu$ is 
decreased and the simulation is repeated. We carefully checked that 
$\nu_M$ did not depend 
on the system size, by using systems with $L$ ranging from 1024 to 
4096. Nevertheless, the present simulations did not show any significant 
deviation from linearity in the dependence of $\nu_M$ on $c_3$, and 
our best fit is
\be
  \nu_M \simeq  -(0.74\pm 0.01) c_3 +0.57 \pm 0.01\;.      \label{lin_fit}
\ee
From this we can estimate the critical value of $c_3$ to be 
$c_3^* = 0.76 \pm 0.03$. Notice that the linearity of $\nu_M$ agrees 
with a mean field description, since the critical exponent for the 
order parameter in cases without spontaneous symmetry breaking is 1.

Previous estimates of $c_3^*$ (for the same value of $c_1$) were all 
made by approaching the transition from the DT side. The defect density 
$\delta_D$ was measured in \cite{shra,green} by counting how often the 
total phase difference $\delta\psi(t) = \psi(x+L,t)-\psi(x,t)$ changed 
when going from $t$ to $t+\tau$. Assuming a scaling law $\delta_D\sim
(c_3 - c_3^*)^\alpha$, the authors of \cite{shra} found $c_3^* \simeq 
0.77$ and $\alpha \simeq 2$. In contrast, $c_3^* = 0.74$ and $\alpha 
\simeq 6.8$ was found in \cite{green}. Since the latter used much higher 
statistics (iteration times up to $10^7$ units, $L=4096$ against 
$L=1024$ in \cite{shra}) one might consider it as more significant. 
But the very large value of $\alpha$ suggests that the data of \cite{green}
presumably do not follow a power law, and the extrapolation to $\delta_D
=0$ is doubtful. Indeed, another extrapolation was also tried in 
\cite{green} which gave $c_3^*=0.70$. Thus our result quoted above is 
in good agreement with previous estimates, in view of the latter's 
uncertainties. 

To obtain an independent estimate of $c_3^*$ and of an eventual scaling 
exponent governing the approach from the DT side, we performed simulations
(with $\nu=0$) for several values $c_3 \geq c_3^*$. In these simulations we 
did not measure actual defects but {\it near} defects. More precisely, 
we made histograms of $P(\rho)$ similar to the one shown in Fig.~2. A 
number of such histograms are shown in Fig.~4 (actually, the quantity 
shown in Fig.~4 is $P(\rho)/(2\pi\rho)$, i.e. the 2-dimensional density). 
We see much wider 
distributions than in Fig.~2, reflecting again the higher degree of 
chaoticity for $\nu=0$. From these histograms we can estimate the 
probabilities
\be
    w(\rho) = \int_0^\rho dx P(x)
\ee
to have an amplitude $<\rho$. We did this for several values of $\rho$, and 
plotted the results against $c_3 - c_3^*$ in a log-log plot (see Fig.~5). 
We see a decent scaling law if $c_3^* =0.734 \pm 0.007$ (which we take 
as our best estimate of $c_3^*$), with an exponent $\simeq 6.3$. This 
large exponent might however suggest that the data should not be described 
by a power law at all \cite{green}. In that case, the true value of 
$c_3^*$ could be considerably smaller and close to the ``alternative'' 
value 0.70 found in \cite{green}. Thus the agreement with previous 
simulations and with Eq.~(\ref{lin_fit}) is as good as can be expected 
in view of these uncertainties.

In Fig.~4 it is seen that $\rho_{min}$ (the lower limit of 
the support of $P(\rho)$) does not approach zero continuously as 
$c_3 \to c_3^*-\epsilon$. Instead, as $c_3$ approaches $c_3^*$ from 
below, it seems that $\rho_{min}$ jumps abruptly to 0. On the other 
hand, the 2-dimensional density $P(\rho)/(2\pi\rho)$ seems to be flat 
at $\rho=0$, for all $c_3>c_3^*$. As a consequence, the distribution of 
local phase gradients $\nu_{loc}$ should decay as $P(\nu_{loc}) \sim 
\nu_{loc}^{-3}$ (here we assume that $\nu_{loc}\sim 1/\rho$ for large 
$\nu_{loc}$), in qualitative agreement with observations in \cite{shra}. 
As pointed out in \cite{shra,green}, the creation of a defect can be 
similar to a noise-induced widening of an attractor. This would 
naturally lead to a defect density which decreases exponentially with 
$c_3$. But we believe that the phenomenon looks more like an (interior)
crisis \cite{grebogi}. It is known that crises
can lead to power laws with very large exponents \cite{yorke} or even 
to characteristic time scales which 
increase faster than any (inverse) power of the distance from the 
critical point \cite{grebogi2}. But it seems
that little is known about crises in spatially extended systems. The 
simplest possibility would be that for each pair $(c_1,\nu)$ there is 
a critical circle $|A|=\rho^*$. Once this circle is 
penetrated, the orbit can come arbitrarily close to the origin, and a 
defect can develop. Unfortunately, we do not see any hint for such a 
circle in our data. The next simple scenario would be that there is 
a rotationally and translationally invariant set of tori 
$\rho=\rho^*(x,\psi)$ (each of which individually is not symmetric), 
such that a defect can appear as soon as the solution is inside any one 
of these tori. We see no way how to test this scenario.

\section{Lyapunov Analysis and Stable Solutions}

To measure the degree of ``chaoticity'' of the solutions with $\nu \ne 0$ 
in the PT regime, we measured the maximal Lyapunov exponent $\lambda$.
We did this for several values of $c_3$ and for several $\nu$-values 
in the interval $[0,\nu_M]$. Our data are shown in Fig.~6, for 
$c_3=0.35,0.40,0.60,$ and 0.70. They were obtained from single trajectories 
spanning between $1.5\times10^5$ and $6\times10^5$ time units. This 
time systematic differences were seen between different initial conditions 
(we used again Eqs.~(\ref{in1}) and (\ref{in2})), indicating broken
ergodicity even for fixed $\nu$. Thus there seem to be several coexisting 
attractors, with slightly different values of $\lambda$. This 
complicates the situation of course, and it might be responsible for 
the slight non-monotonicities seen in Fig.~6. Apart from these, the 
most evident feature seen in Fig.~6 is the strong decrease of $\lambda$ 
with $\nu$. Indeed, it seems that in general $\lambda\to 0$ for 
$\nu \to \nu_M$. This decrease is seen even more clearly if averages 
are taken over several initial conditions, as is done in Fig.~7 for 
$c_3=0.5$. In this figure, $\lambda$ was averaged over 15 - 35 
trajectories for each value of $\nu$, with $t=10^5$ for each trajectory.

The progressive ordering of the states with increasing $\nu$ can be 
directly seen when plotting snapshots of individual solutions. A sequence 
of such snapshots is shown in figures 8 (amplitudes) and 9 (phases). 
While the configurations seem chaotic for small $\nu$, they are more 
and more regular for larger $\nu$, to become finally periodic for $\nu
\approx \nu_M$. While it is not surprising that the measured $\lambda$ 
was zero for Figs.~8e and 9e, it {\it is} surprising that the 
same is also true for Figs.~8d and 9d. Roughly, the latter solution can 
be described as a sequence of periodic ``patches'' (each similar to 
the solution shown in Figs.~8e and 9e), interrupted by ``faults'' in 
which the amplitude is nearly constant and the phase decreases linearly.

We have observed such non-chaotic solutions for all $c_3 \le 0.5$
when $\nu \to \nu_M$. They can all be expressed in the form
\be
    A(x,t)=h(x-vt){\rm e}^{i(\nu x - \omega t)}           \label{wave1}
\ee
where $h(\xi)=\rho(\xi)\,e^{i\psi_0(\xi)}$ is in general complex and 
periodic with period $L$. Notice that also its phase is periodic, so that 
$\nu$ in Eq.~(\ref{wave1}) is the total average phase gradient. We found 
both solutions which are periodic with period $L$ (forced upon them by 
the periodic boundary condition), and solutions with smaller period 
(as, e.g. in Fig.~8e and 9e). It is natural to assume that the former 
would lead to spatially aperiodic solutions for $L\to\infty$. The 
periodic solutions can be considered as Bloch waves where the periodicity 
is however self generated and not imposed by some external potential.

Correspondingly, amplitude and phase can be expressed as
\be
   \rho(x,t)=\rho(\xi) \quad ; \quad \psi(x,t)=\psi_0(\xi)-\omega t+\nu x\;.
                                                         \label{wave2}
\ee
where $\xi = x-vt$.
While the motion of the amplitude is a simple shift with velocity 
$v$, the evolution of the phase is more complicated. At fixed $\xi$, 
i.e. in a coordinate frame moving with velocity $v$, the phase increases 
linearly with $\partial \psi/\partial t|_\xi = v\nu- \omega$. But there 
is no fixed phase velocity, i.e. there is no frame in which the phase 
is strictly constant. It fluctuates around a constant value in a frame 
moving with velocity $\langle v_{ph} \rangle = \omega/\nu$, which is 
thus the {\it average} phase velocity. Typically, we found 
that $v$ and $\langle v_{ph}\rangle$ are not commensurate. Thus solutions 
of the type of Eq.~(\ref{wave1}) display a novel type of quasiperiodic 
motion. Similar waves have been found previously in \cite{janiaud,baleari}.

While time dependent simulations are needed to verify the stability of 
these solutions, their {\it existence} can be checked much more easily. 
Indeed, they are obtained by solving an ordinary differential equation 
for $h(\xi)$, 
\be
    (1+ic_1)h_{\xi\xi}+[2i\nu(1+ic_1) + v] h_\xi + 
           [1-\nu^2(1+i c_1)+i\omega] h-(1-ic_3)|h|^2 h = 0
                                                            \label{ode}
\ee

We should notice that for $c_3 \ge 0.6$ the picture changes slightly: 
$\lambda$ still decreases with $\nu$, but non-chaotic solutions are no 
longer observed.

Another way to simulate regular solutions with periodic $h(\xi)$ is to 
use small systems where $L$ is equal to its period. To obtain arbitrary 
values of $\nu$, we have to use in this case {\it twisted} boundary 
conditions $\rho(x+L)=\rho(x), \psi(x+L)=\psi(x)+\theta$ with a twist 
$\theta = \nu L$. By means of such simulations we checked again the 
existence of solutions (\ref{wave1}). We checked also that there was 
an upper limit on $\nu$ which agreed roughly with $\nu_M$, though its 
precise value depended on $L$, as one might have expected: depending 
on its precise value, an integer number of waves will fit into $L$ 
with more or less ease.

\section{A modified Kuramoto-Sivashinsky equation}

The behavior of $\lambda$ as a function of $\nu$ and the origin of 
the observed non-chaotic solutions can be explained in the framework of 
the modified Kuramoto-Sivashinsky equation (MKSE) derived in 
\cite{saga,saga2}. For completeness, let us recall the main assumptions 
in deriving the Kuramoto-Sivashinsky equation (KSE) \cite{kura,shiva} 
and its modification.

It is well known that just above the BF line the amplitude is nearly 
constant, and its dynamics is essentially ruled by the phase behavior
\cite{kura_book}. More precisely, the KSE is obtained by making two main
assumptions: first, $A(x,t)$ is obtained by adding a small perturbation 
to the spatially constant solution, $A(x,t) = e^{-i\Omega_0 t}+u(x,t)$; 
and secondly, $u(x,t)$ is a ``slaved'' function of $\psi(x,t)$ and its 
spatial derivatives. 

In deriving the MKSE we just replace the first assumption by making a 
perturbation around the plane wave solution given in Eq.~(\ref{plane}).
After a tedious and non-trivial calculation \cite{saga} one finds that 
the amplitude is given by 
\be
    \rho(x,t)=\sqrt{(1-q(x,t))^2+(c_3 q(x,t))^2}     \label{saga_amp}
\ee
with $q(x,t)=[c_1 \partial_x^2 \psi(x,t) + (\partial_x \psi(x,t))^2]/2$, 
and the phase satisfies 
\be
    {\dot \psi} + \Omega_2^{(1)} \psi_{xx} 
                + \Omega_2^{(2)} (\psi_x)^2
                + \Omega_4^{(1)} \psi_{xxxx}
                + \Omega_4^{(2)} \psi_x \psi_{xxx} = 0      \label{saga}
\ee
where $\Omega_2^{(1)}= c_1 c_3-1$, $\Omega_2^{(2)}=(c_1+c_3)$,
$\Omega_4^{(1)}=c_1^2 (1+c_3^2)/2$ and $\Omega_4^{(2)}=2 c_1 (1+c_3^2)$.
Notice that the latter is just the KSE except for the last term. 

In order to verify if Eqs.~(\ref{saga}) and (\ref{saga_amp}) well
approximate the dynamics of the CGLE in the PT regime, two checks have 
been performed. As a first test, we have evaluated $\rho(x,t)$ by means 
of Eq.~(\ref{saga_amp}) in the whole PT regime and for several values of 
$\nu$. A comparison with the amplitudes obtained directly from the 
integration always shows satisfactory results, see Fig.~10. Notice that this 
is even true for the extremal values $c_3 = 0.4$ and $0.7$, i.e. for two  
parameter values which are near to the BF line and to the line $L_1$. 

As a second check, we have derived from Eq.~(\ref{saga}) expressions 
for the amplitude velocity $v$ and the frequency $\omega$ associated to 
the non-chaotic solutions (\ref{wave1}). We first observe that the 
ansatz (\ref{wave1}) leads to 
\be
    \omega = - {\dot \psi}(x,t) -v \psi_0'(\xi) \quad ; \quad 
    \partial_x \phi(x,t) = \psi_0' (\xi) \;\;,            \label{deriv}
\ee
where primes indicate derivatives with respect to $\xi = x-vt$.
We then substitute this into (\ref{saga}) and integrate over all $x$. Due 
to the periodic boundary condition some terms drop out and others can be 
simplified, leaving us with
\be
    \omega = \Omega_0 + \Omega_2^{(2)} \langle (\psi_0')^2 \rangle
       -\Omega_4^{(2)} \langle (\psi_0'')^2 \rangle \;.      \label{om}
\end{equation}  
where brackets indicate spatial averages, $\langle \cdot \rangle = 
L^{-1}  \int_0^L dx \; \cdot \;$ (a similar result was derived in 
\cite{saga2}, but with the last term neglected). To obtain an expression 
for $v$ we multiply both sides of (\ref{saga}) by $\psi_0'(\xi)$ 
before averaging, arriving at
\be
   v = {\langle \Omega_2^{(2)}[(\psi_0')^3+2\nu(\psi_0')^2] + 
                \Omega_4^{(2)}[(\psi_0')^2\psi_0'''-\nu(\psi_0'')^2]\rangle
          \over  \langle(\psi_0')^2 \rangle}\;.                \label{vu}
\ee
Numerical tests for several parameter values are shown in Table 1. In 
all cases the agreement between the measured values and the right hand 
sides of Eqs.~(\ref{om}) and (\ref{vu}) is very good.

Having verified that Eq.~(\ref{saga}) gives a good approximation for the
phase dynamics of the CGLE in the whole PT regime, we can now relate 
our numerical findings to several other observations in the literature.
This is again based on an observation by Sakaguchi \cite{saga2} who noticed
that Eq.~(\ref{saga}), once $\psi_x$ is approximated by its average value 
$\nu$ in the last term, can be rewritten as an equation for $s=\psi_x$:
\be
   {\dot s} + \Omega_2^{(1)} s_{xx} + 2\Omega_2^{(2)} ss_x
            + \Omega_4^{(1)} s_{xxxx} + \Omega_3 s_{xxx} = 0 \;.  
                 \label{kawa_eq}
\ee
with $\Omega_3=\nu\Omega_4^{(2)}$.
This is exactly the so-called Kawahara equation \cite{kawa,kawa2} 
which has been studied numerically \cite{kawa} and theoretically 
\cite{kawa2,chang}. Periodic wave trains formed by pulse-like structures 
have been found in this equation. It has been argued \cite{kawa}
that they are due to the dispersion term $\propto s_{xxx}$, a conclusion 
which was confirmed in a careful analysis by Chang {\it et al.} 
\cite{chang}. These studies confirm also the coexistence of chaotic and
non-chaotic solutions (depending on the initial conditions). Chaotic 
attractors dominate for small values of the dispersion constant 
$\Omega_3$, while for increasing $\Omega_3$ periodic attractors prevail. 
Finally, above a threshold value, only non-chaotic solutions are found.
Since $\Omega_3 \propto \nu$, our results about the decrease of $\lambda$ 
with increasing $\nu$ and the appearance of wave train solutions for 
$\nu \to \nu_M$ are fully consistent with the findings of \cite{chang}. 
Finally, the observation of Chang {\it et al.} that stable solutions exist 
only below a certain threshold --- which in our case turns out to be 
$\nu \simeq 0.1$ for $0.35 \le c_3 \le 0.5$ --- explains why no regular 
solutions were found above $c_3 = 0.6$: for $c_3 > 0.6$ the value of 
$\nu_M$ is below 0.1.

\section{Concluding Remarks}

In this paper a new integration scheme has been introduced which is 
faster than the usual pseudo-spectral code at comparable accuracy.
It has been successfully employed to study the CGLE in the PT regime for 
large systems ($L > 1024$) and long integration times ($t > 10^5$).
Although we have not applied it to other systems than the CGLE, we 
believe that it should be useful also for the nonlinear Schr\"odinger 
equation and for chemical reaction-diffusion equations.
 
Particular attention was paid to solutions with non-zero average phase 
gradient $\nu$. For fixed control parameters we always found a maximal 
conserved phase gradient $\nu_M$. It can be used as an order parameter to 
characterize the PT regime and to describe the transition from phase to 
defect turbulence: at the transition, $\nu_M$ vanishes with the mean field 
exponent. Another order parameter with non-zero values in the DT regime 
gave the same transition point but another exponent which is more similar 
to exponents found in previous papers. In contrast, the minimal amplitude
values reached in the PT regime do not seem to be useful order parameters, 
as they are discontinuous at the transition.

For small values of $\nu$, the majority of the observed solutions are 
chaotic. But the maximal Lyapunov exponent typically decreases with $\nu$, 
and stable solutions are more frequent than chaotic ones near $\nu_M$.
This can be understood by assuming that the phase dynamics is ruled by a 
modified Kuramoto-Sivashinski equation. 

The regular solutions can be either spatially periodic or aperiodic. 
In both cases they consist of traveling waves whose amplitude and phase 
velocities are typically incommensurate. These solutions can be considered 
as generalized Bloch waves with self generated (periodic or aperiodic) 
modulation.
For spatially periodic regular solutions these results were confirmed by 
simulations of small systems with twisted boundary conditions (i.e., 
periodic amplitudes but $\psi(x+L) = \psi(x)+\theta$). 

During the final write-up of this paper, we became aware of a paper
by Montagne, Hernandez-Garcia and San Miguel \cite{baleari} where results
similar to those reported in Sections III have been found. The analysis
reported in \cite{baleari} confirms that $\nu_M$ can be used as an order 
parameter in the whole $(c_1,c_3)$-plane of the CGLE. 

\acknowledgments

We want to thank M. San Miguel for sending us reference 
\cite{baleari} prior to publication, and to A. Politi for very 
helpful discussions. One of us (A.T.) is supported by the European 
Community through grant no ERBCHBICT941569, and by the Cooperativa 
Fontenuova. This work was also supported by the DFG through the 
Graduiertenkolleg ``Feld\-theoretische und numerische Methoden in der 
Elementarteilchen- und Statisti\-schen Physik'', and through SFB 237.

\begin{figure}
\noindent
{\bf Fig.1}: Mean square errors due to space discretization and 
truncation of the convolution for our algorithm (symbols), and due 
to space discretization for a time splitting pseudo-spectral code (solid 
lines). The asteriscs show the errors due to the time discretization
(they are the same for both algorithms). All the data refer to $c_1=3.5$, 
$c_3=0.9$, $\tau = 0.05$ and $N=1024$. The results for our algorithm
are reported for several values of the number of convolution channels $N$. 
The simulations were done with double precision, whence the round-off 
errors are at $\sim 10^{-15}$.
\end{figure}

\begin{figure}
\noindent
{\bf Fig.2}: Probability distribution $P(\rho)$ for $c_3=0.6$,$c_1=3.5$
and for three different values of $\nu$: 0.015 (solid line); 0.046 (dotted 
line) and 0.098 (dashed line).
\end{figure}

\begin{figure}
\noindent
{\bf Fig.3}: Maximal phase gradient $\nu_M$ as a function of the parameter
$c_3$ (asterisks). The solid line represent a linear fit for the $\nu_M$ data. 
\end{figure}

\begin{figure}
\noindent
{\bf Fig.4}: Probability distributions $P(\rho)$ similar to those shown 
in Fig.~2, but for $\nu=0$, divided by $2\pi\rho$, and shown on a logarithmic 
scale. For all curves $c_1=3.5$. From bottom to top, the values of $c_3$ are 
0.74, 0.746, 0.75, 0.752, 0.755, 0.758, 0.761, 0.765, 0.77, 0.783, 0.8, 0.82,
0.85, and 0.9. 
Except possibly for the bottom most, all curves are in the DT regime. The 
simulations closest to $c_3^*$ were done with $L=4096$ and extended over 
$>2\times10^5$ time units.
\end{figure}

\begin{figure}
\noindent
{\bf Fig.5}: Log-log plot showing the probabilities to find an amplitude 
$<\rho$ against $c_3-0.734$. Each curve corresponds to a fixed value of 
$\rho$, with $\rho=m/20$ for the $m$-th curve from the bottom.
\end{figure}

\begin{figure}
\noindent
{\bf Fig.6}: Maximal Lyapunov exponents $\lambda$ versus the phase 
gradient $\nu$ for $c_3=0.35$ (a); $c_3=0.40$ (b); $c_3=0.6$ (c) and 
$c_3=0.7$ (d). Two different sets of initial conditions are considered: 
(\ref{in1}) (asterisks) and (\ref{in2}) (circles). The data have been 
obtained for system sizes $L=1024$ and $L=2048$ and for integration 
times $t =$ 150,000 - 600,000.  
\end{figure}

\begin{figure}
\noindent
{\bf Fig.7}: Maximal Lyapunov exponents $<\lambda>$ for $c_3=0.5$, 
plotted against $\nu$. Each value has been obtained averaging over
many initial conditions and over an integration time $t=100,000$ for 
each trajectory. The considered system size is $L=1024$. 
\end{figure}

\begin{figure}
\noindent
{\bf Fig.8}: Amplitude $\rho(x)$ for $c_3=0.5$ and for various values of 
the phase gradient $\nu$: 0 (a); 0.01 (b); 0.04 (c); 0.09 (d) and 0.18 (e).
\end{figure}

\begin{figure}
\noindent
{\bf Fig.9}: Same as Fig.~8, but for the phase $\phi(x)$.
\end{figure}

\begin{figure}
{\bf Fig.10}: Amplitudes $\rho$ from the direct simulation of the
CGLE (solid line) and obtained through Eq.~(\ref{saga_amp}) 
(dashed line):
a) $c_3=0.7$ and $\nu=0.06$;
b) $c_3=0.7$ and $\nu=0$;
c) $c_3=0.4$ and $\nu=0.26$;
d) $c_3=0.4$ and $\nu=0.03$.
\end{figure}

\begin{table}
\caption[tabone]{
Amplitude velocity $v$ and the frequency $\omega$ of non-chaotic solutions 
(see Eq.~(\protect{\ref{wave1}})) for several values of $c_3$ and $\nu$. 
The first numbers in columns 3 and 4 were obtained from 
Eqs.~(\protect{\ref{om}}) and (\protect{\ref{vu}}), while the directly 
measured values are given in parentheses.
}

\vskip 1 truecm

\begin{tabular}{rrrr}
\hfil $c_3$ \hfil \hfil   &
\hfil $\nu$ \hfil \hfil & \hfil $\omega$ \hfil & \hfil $v$ \hfil 
\hfil \\ 
\hline
\hline
0.35  &  0.092 &  0.317 (0.317) & 0.723 (0.726) \\
0.35  &  0.184 &  0.218 (0.218) & 1.454 (1.458) \\
0.35  &  0.306 &  -0.021 (-0.021) & 2.563 (2.569) \\
0.40  &  0.092 &  0.365 (0.365) & 0.731 (0.735) \\
0.40  &  0.153 &  0.305 (0.305) & 1.236 (1.241) \\
0.40  &  0.282 &  0.074 (0.070) & 2.372 (2.337) \\
0.50  &  0.123 &  0.428 (0.427) & 1.256 (1.241) \\
0.50  &  0.184 &  0.355 (0.341) & 1.592 (1.650) \\
0.50  &  0.245 &  0.229 (0.223) & 1.933 (1.984) \\
\end{tabular}
\end{table}

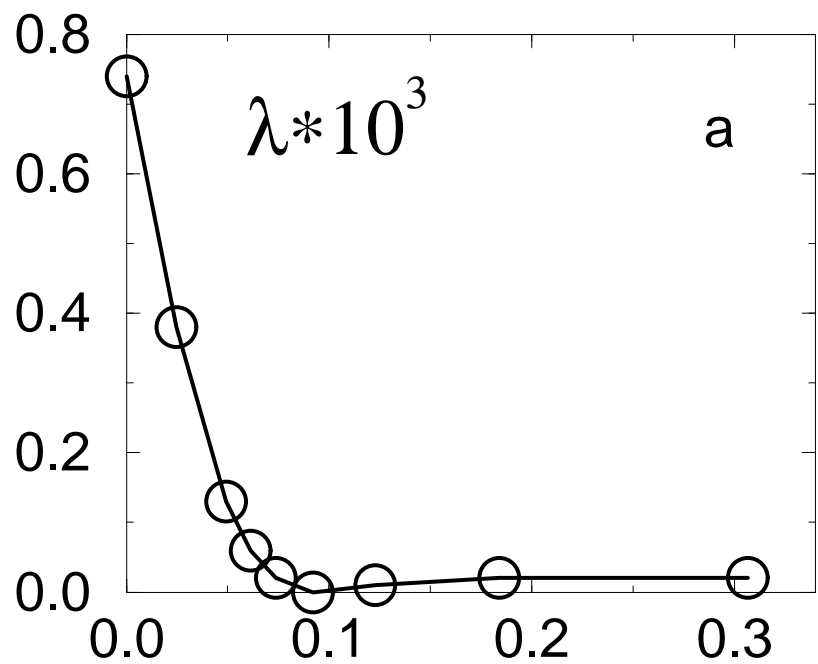
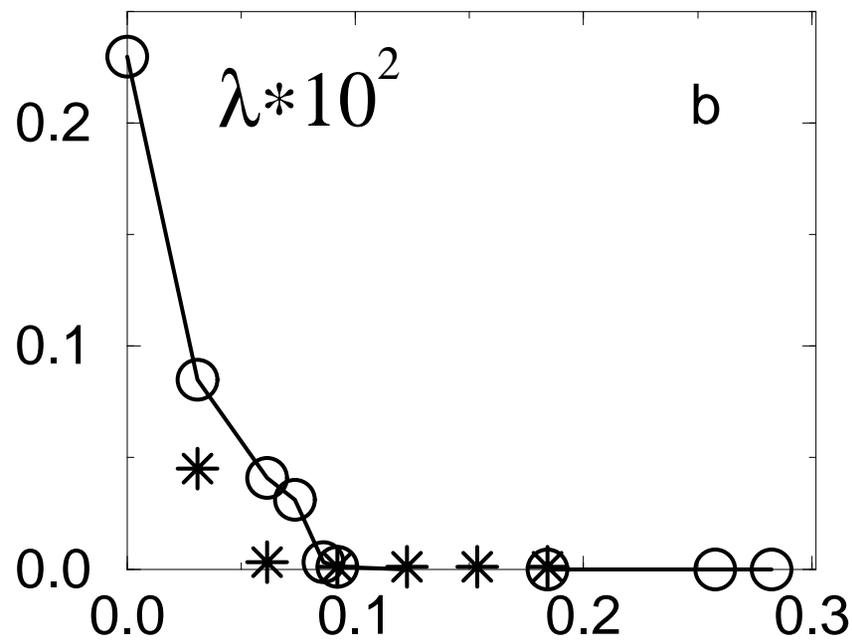
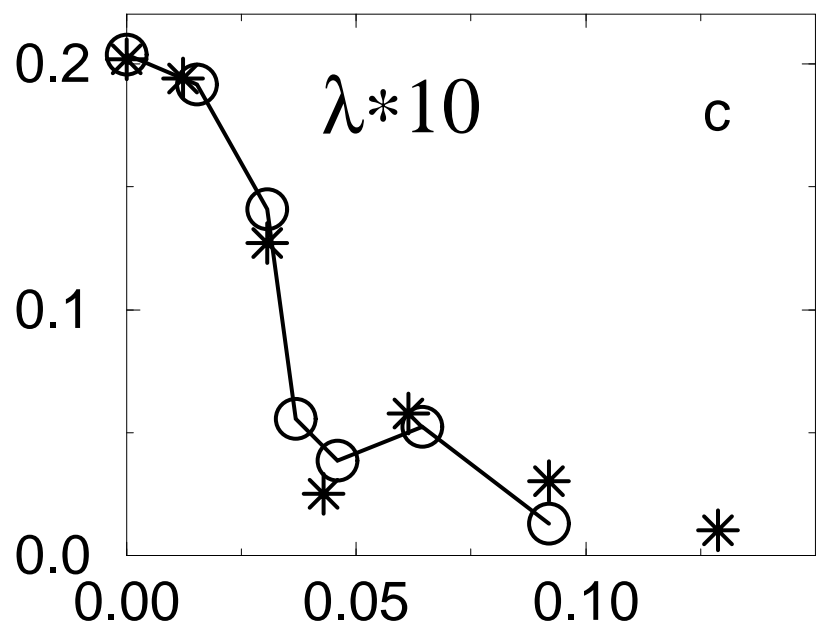
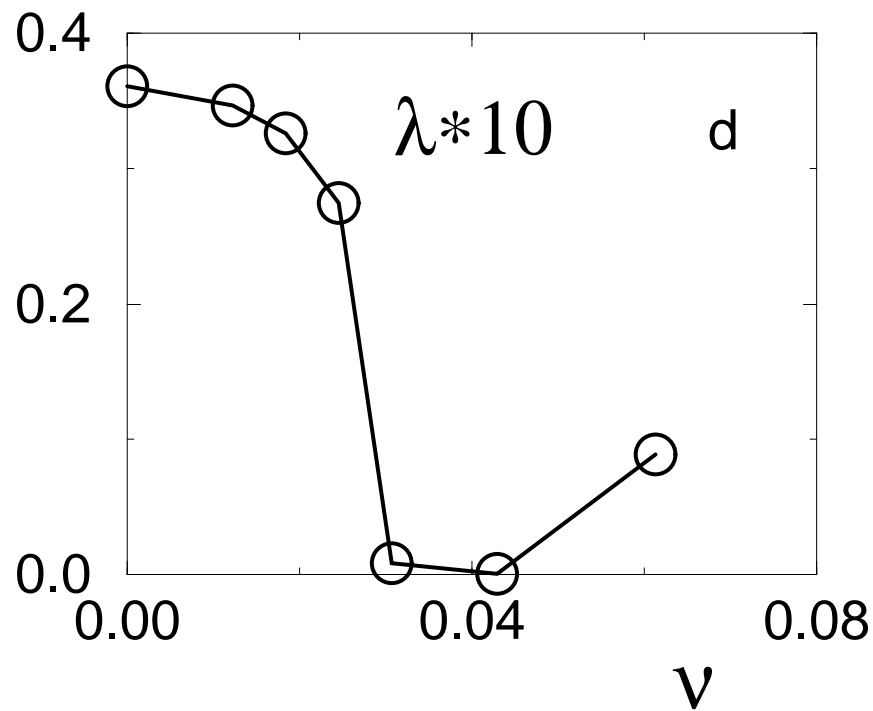
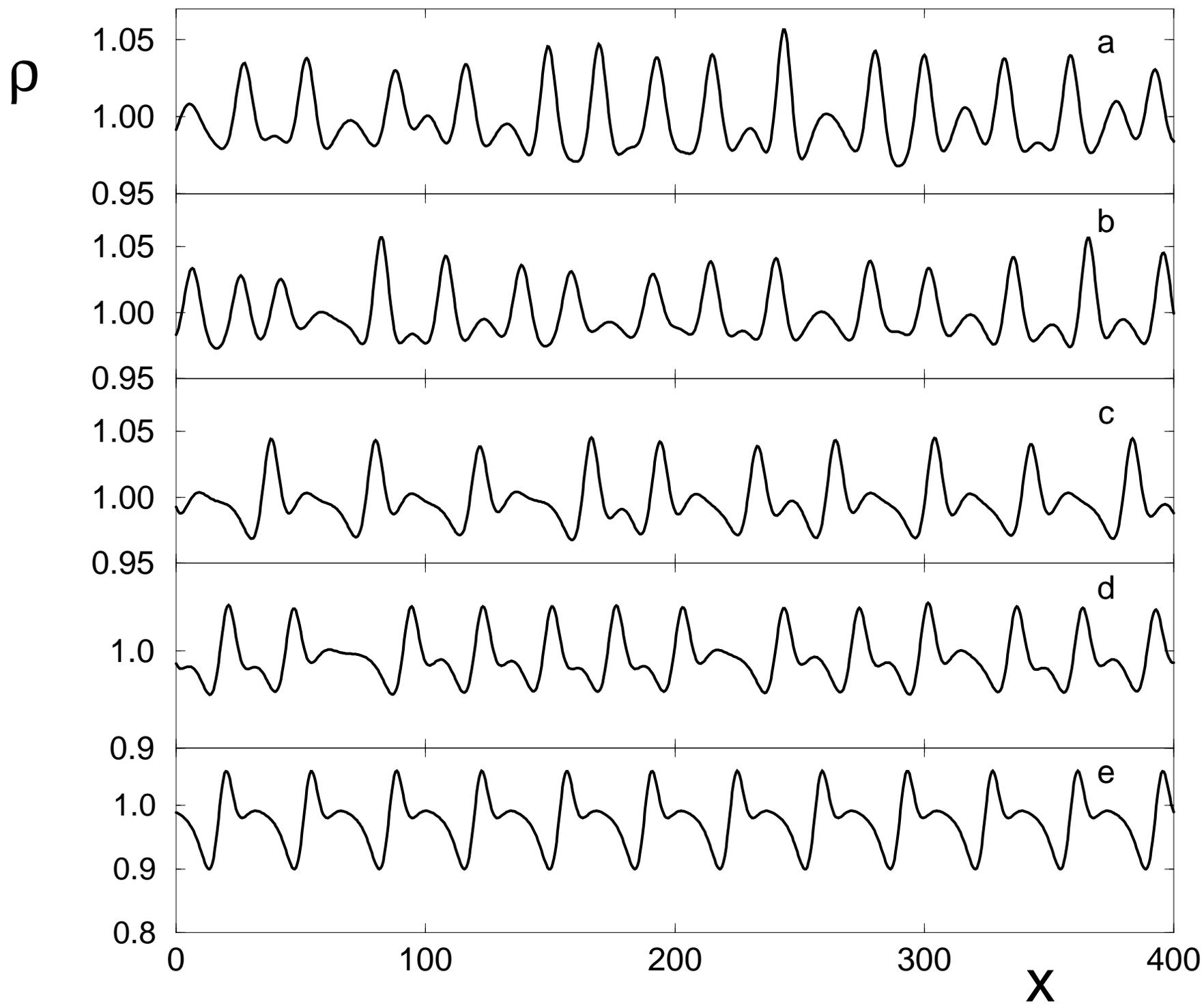
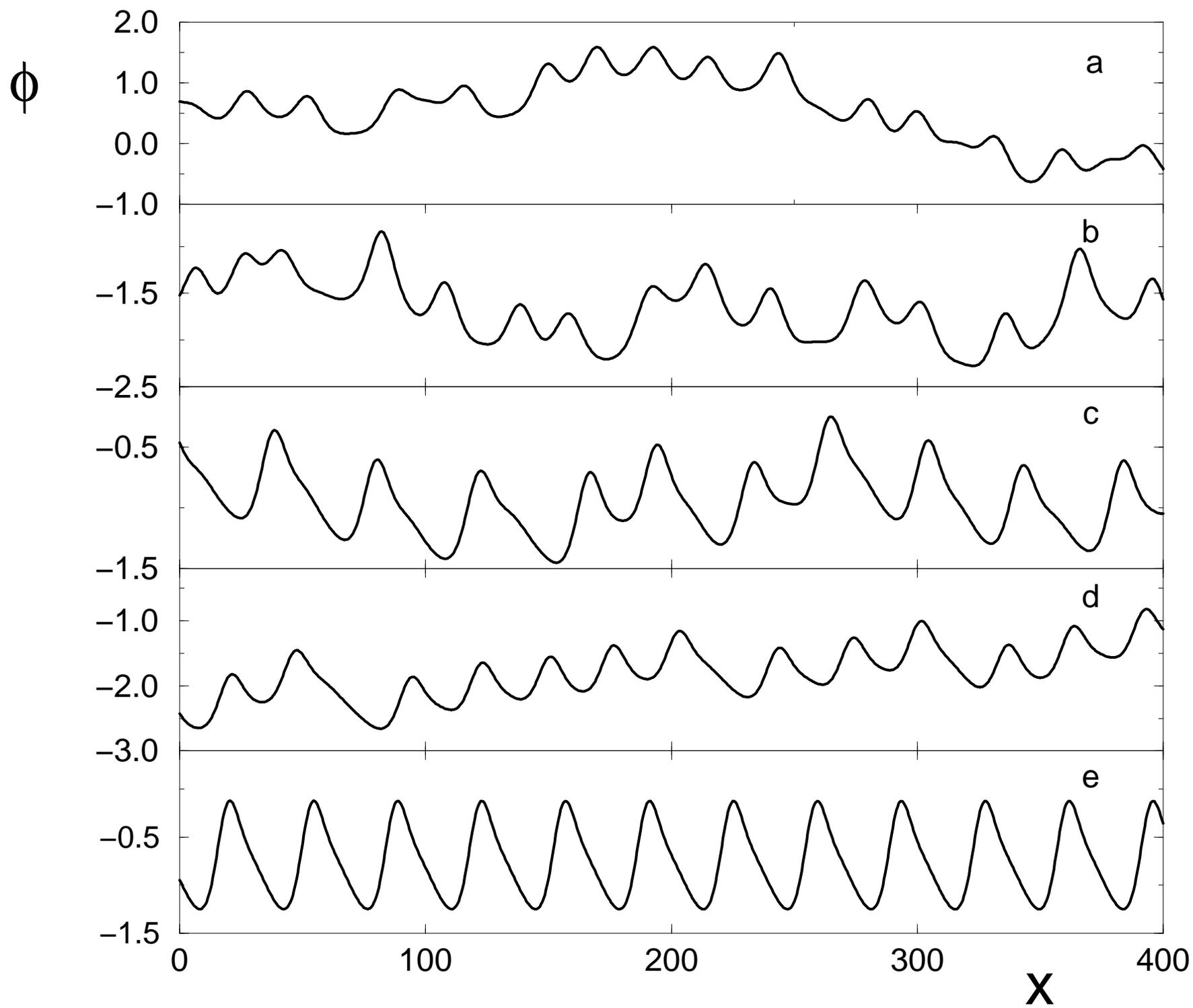
\end{document}